\documentclass[12pt]{emulateapj}

\renewcommand{\deg}{\mbox{$^{\circ}$}}

\def\deg{\ifmmode^\circ\else$^\circ$\fi}    

\def\hper{\ifmmode \rlap.^{h}\else $\rlap{.}^h$\fi} 
\def\sper{\ifmmode \rlap.^{s}\else $\rlap{.}^s$\fi}    

\def\deg{${}^\circ$}

\def\today{\number\year\space \ifcase\month\or
  January\or February\or March\or April\or May\or June\or
  July\or August\or September\or October\or November\or December\fi
  \space\number\day}
\def\now{\number\year\space \ifcase\month\or
  January\or February\or March\or April\or May\or June\or
  July\or August\or September\or October\or November\or December\fi
  \space\number\day .\number\time}

\shorttitle{On the stellar content of IC10} 
\shortauthors{Sanna et al.}

\begin{document}
\title{On the stellar content of the starburst galaxy IC10\altaffilmark{1}}

\author{
N.\ Sanna\altaffilmark{2,3},
G.\ Bono\altaffilmark{2,3},
P.\ B.\ Stetson\altaffilmark{4},
A.\ Pietrinferni\altaffilmark{5},
M.\ Monelli\altaffilmark{6},
S.\ Cassisi\altaffilmark{5}, 
R.\ Buonanno\altaffilmark{2},
E.\ Sabbi\altaffilmark{7}
F.\ Caputo\altaffilmark{3},
M.\ Castellani\altaffilmark{3},
C.\ E.\ Corsi\altaffilmark{3},
S.\ Degl'Innocenti\altaffilmark{8,9}
I.\ Drozdovsky\altaffilmark{6},
I.\ Ferraro\altaffilmark{3},
G.\ Iannicola\altaffilmark{3},
M.\ Nonino\altaffilmark{10}, 
P.\ G.\ Prada Moroni\altaffilmark{8,9},
L.\ Pulone\altaffilmark{3},  
M.\ Romaniello\altaffilmark{11}, and  
A.\ R. \ Walker\altaffilmark{12}
}

\altaffiltext{1}
   {Based on observations collected with the ACS and the WFPC2 
on board of the HST.}
  \altaffiltext{2}{Univ. Roma ToV, via della Ricerca Scientifica 1, 00133 Rome, Italy; Nicoletta.Sanna@roma2.infn.it}
 \altaffiltext{3}{INAF--OAR, via Frascati 33, Monte Porzio Catone, Rome, Italy}
 \altaffiltext{4}{DAO--HIA, NRC, 5071 West Saanich Road, Victoria, BC V9E 2E7, Canada}
 \altaffiltext{5}{INAF--OACTe, via M. Maggini, 64100 Teramo, Italy}
 \altaffiltext{6}{IAC, Calle Via Lactea, E38200 La Laguna, Tenerife, Spain}
 \altaffiltext{7}{STScI, 3700 San Martin Drive, Baltimore, MD, 21218, USA}
 \altaffiltext{8}{Univ. Pisa, Largo B. Pontecorvo 2, 56127 Pisa, Italy}
 \altaffiltext{9}{INFN, Sez. Pisa, via E. Fermi 2, 56127 Pisa, Italy}
 \altaffiltext{10}{INAF--OAT, via G.B. Tiepolo 11, 40131 Trieste, Italy}
 \altaffiltext{11}{ESO, Karl-Schwarzschild-Str. 2, 85748 Garching bei Munchen, Germany}
 \altaffiltext{12}{CTIAO--NOAO, Casilla 603, La Serena, Chile}

\date{\centering drafted \today\ / Received / Accepted }

\begin{abstract}
We investigate the stellar content of the starburst dwarf galaxy IC10 
using accurate and deep optical data collected with the Advanced Camera 
for Surveys and with the Wide Field Planetary Camera 2 on board the 
Hubble Space Telescope. The comparison 
between theory and observations indicates a clear 
%
change in age distribution  
when moving from the center toward the external regions. Moreover, 
empirical calibrators and evolutionary predictions suggest the 
presence of a spread in heavy element abundance of the order of 
one-half dex. The comparison between old and intermediate-age core 
He-burning models with a well defined overdensity in the 
color-magnitude diagram indicates the presence of both 
intermediate-age, red clump stars and of old, red horizontal 
branch stars.         
\end{abstract}


\maketitle

\section{Introduction}

Dwarf galaxies are ubiquitous stellar systems outnumbering giant systems  
in the Local Group (LG, Mateo 1998), in the Local Volume ($d\le 10$ Mpc, 
Vaduvescu \& McCall \ 2008), and in the nearby Universe 
(Popesso et al.\ 2006; Milne et al.\ 2007). Recent 
evidence indicates that the Local Group includes at least 62 dwarfs 
and among them 
$26\pm5$\% are dwarf irregulars (dIrrs, Grebel 2003; McConnachie et al.\ 2008). 
However, we still lack firm criteria discriminating between dIrrs and 
Blue Compact Galaxies (BCDs). According to Thuan (1985) and to 
van den Bergh (2000) the BCDs are dIrrs that are experiencing a significant 
burst of star formation. On the other hand, Richer \& McCall (1995) found 
that the metal abundance of BCDs is more 
similar to dwarf spheroidals than to dIrrs, and Papaderos et al.\ (1996) 
pointed out the lack of an evolutionary link among BCDs, dIrrs and 
dwarf ellipticals (dEs). This key issue is far from being settled, and 
indeed in a recent detailed photometric and spectroscopic investigation 
Vaduvescu \& McCall (2008) suggested that BCDs, dIrrs and dEs define 
the same fundamental plane. 

Dwarf irregulars also play a key role in constraining the impact 
that structural parameters and intrinsic properties have on the evolution 
(initial mass function, star formation history) of complex systems 
(Massey et al.\ 2007). Moreover, they are fundamental laboratories 
for investigating the evolution of massive stars in systems that are 
undergoing significant bursts of star formation (Crowther 2007).      

Among the nearby dIrrs IC10 is an interesting system, since it is one of 
the most massive ($\log M/M_\odot$=8.49, Vaduvescu et al.\ 2007), and  
the comparison between the $H_\alpha$ and the $B$-band luminosity 
indicates that it is experiencing a starburst phase (Hunter et al.\ 1993).  
Moreover, it has been suggested that IC10 harbors a large number of 
young Wolf-Rayet stars (Massey \& Holmes 2002) and intermediate-age 
carbon stars (Demers et al.\ 2004). However, we still lack detailed 
knowledge of the stellar content of IC10. In particular, Vacca 
et al.\ (2007), using deep optical and near-infrared data, proposed
that the isochrone fit to IC10---at fixed distance and metal 
content---would require different reddening values for 
Main Sequence (MS) and Red Giant Branch (RGB) stars.  

In a previous investigation (Sanna et al.\ 2008) we provided a new estimate of
the distance modulus ($\mu$=24.60$\pm 0.15$ mag) based on a new calibration of
the tip of the RGB, and of the reddening (E($F555W$-$F814W$)=1.16$\pm 0.06$ mag)
based on empirical calibrators.  Here we address the galaxy's stellar content. 

\section{Results and discussion}

The photometric catalog we adopt is based on archival optical images 
from the Advanced Camera for Surveys (ACS) and the Wide Field Planetary 
Camera 2 (WFPC2; see top panel of Fig.~1 and Sanna et al. 2008). 
The final catalog includes $\sim 720,000$ stars with at least one measurement
in each of two different bands. The ACS data in the $F555W$ and $F814W$ bands
were placed on the VEGAMAG system following Sirianni et al.\ (2005). 
The $F606W$-band images collected with the ACS were transformed into the 
$F555W$-band using local standards, and the same approach was adopted to 
transform the $F555W$ and the $F814W$ images collected with WFPC2 into 
the corresponding ACS bands.  On average the star-to-star precision of 
the above transformations is better than 0.02~mag (Sanna et al.\ 2009,
in preparation). 
The final catalog was split into two different regions: region {\em C)} 
covers the galaxy center and includes both ACS and WFPC2 data, while 
region {\em E)} lies at a radial distance greater than two arcminutes
and only includes ACS data (see the blue and red polygons 
in the top panel of Fig.~1 and Sanna et al.\ 2008).

\begin{figure*}[!ht]
\begin{center}
\label{fig1}
\includegraphics[height=0.35\textheight,width=0.60\textwidth]{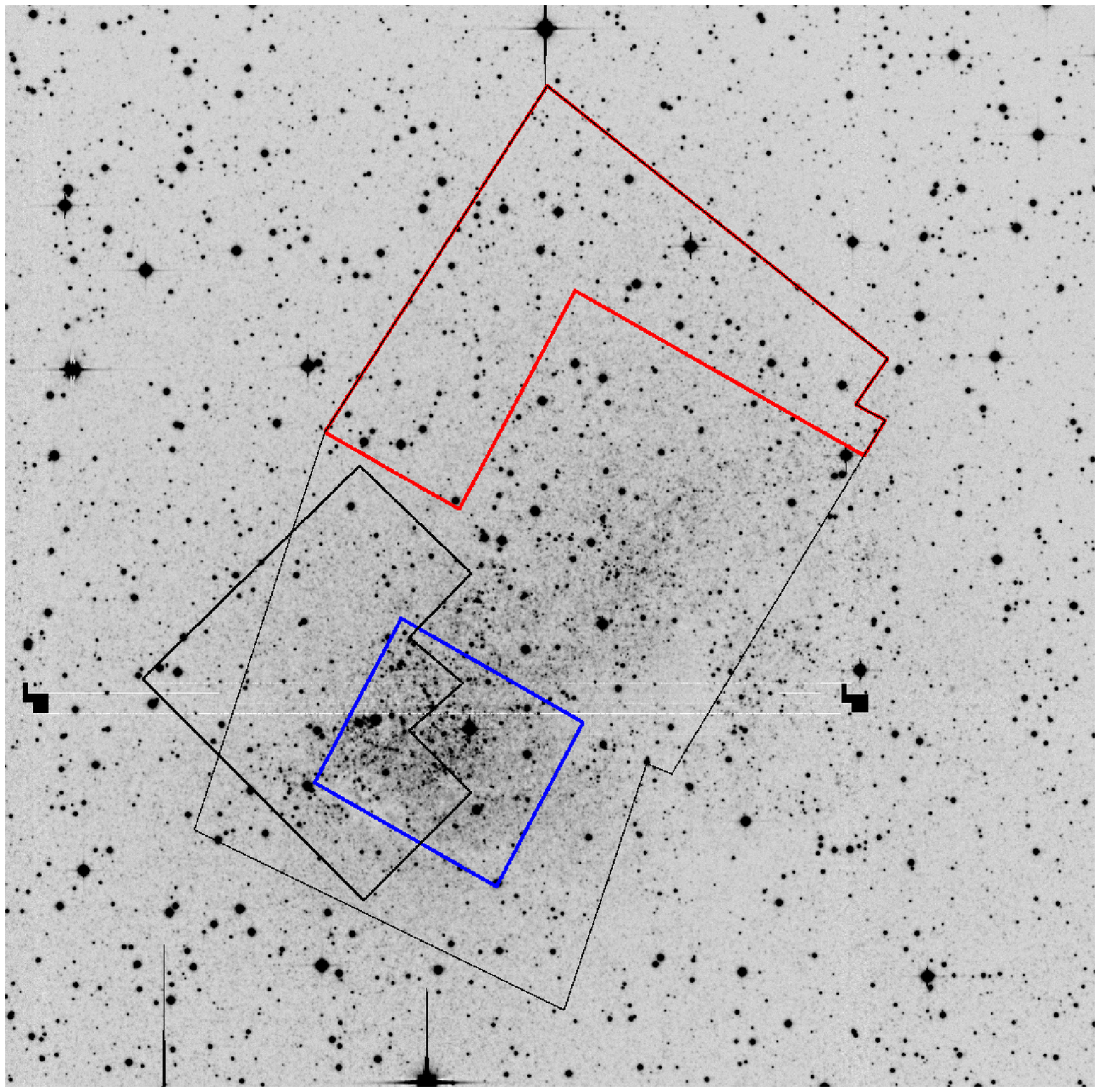}
\includegraphics[height=0.50\textheight,width=0.80\textwidth]{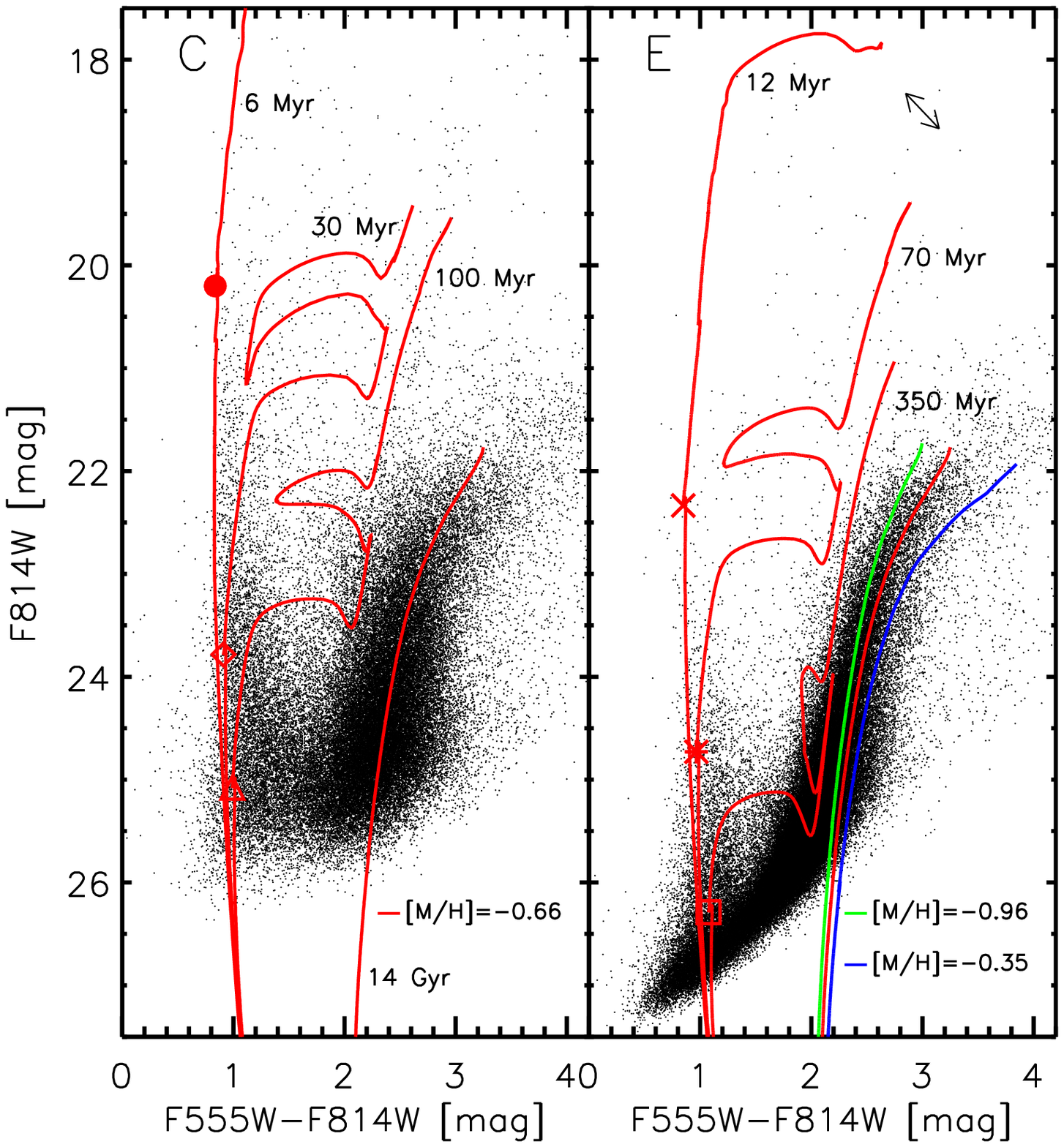}
\vspace*{-0.15truecm}
\caption{Top -- The coverage of the HST data sets collected with 
the ACS and with the WFPC2 (black lines). The blue and red polygons, 
superimposed to IC10, mark fields C) and E), respectively. 
The background is a MegaCam@CFHT image of 6$\times$7 arcmin.
North is up and East is to the left.     
Bottom left -- $F814W$, $F555W$-$F814W$ CMD of IC10 central regions.
Red lines display scaled-solar isochrones (BaSTI) at fixed chemical 
composition and different ages (see labeled values). The circle 
($M(TO)/M_\odot$=30.5), the diamond ($M(TO)/M_\odot$=7.7) and the 
triangle ($M(TO)/M_\odot$=4.3) mark the Turn-Off (TO) of three young 
isochrones. Bottom right -- Same as the left, but for the IC10 external 
regions. 
The cross ($M(TO)/M_\odot$=14.2), the asterisk ($M(TO)/M_\odot$=5.0) and 
the square ($M(TO)/M_\odot$=2.5) mark the TO of three young-intermediate 
age isochrones. The green and the blue lines show two old ($t$=14 Gyr) 
isochrones at different metal contents.  
The arrows in the top right corner show the shift in color and in 
magnitude caused by the possible occurrence of a differential 
reddening of $\pm$10\%.}
\end{center}
\end{figure*}

Spectroscopic estimates of IC10's metallicity, based on HII regions, 
indicate a metal content ($[Fe/H]\sim-0.71\pm0.14$, Lee et al.\ 2003) 
similar to the Small Magellanic Cloud 
(SMC, $[Fe/H]\approx -0.7$, Zaritsky et al.\ 1994 also based on HII regions or 
$[Fe/H]\sim-0.75\pm0.08$, Romaniello et al.\ 2008, based on Cepheids). 
To characterize the stellar content of IC10 we adopted different sets of 
both scaled-solar (young and intermediate ages) and $\alpha$-enhanced 
(old ages) isochrones from the BaSTI 
database\footnote{http://www.oa-teramo.inaf.it/BASTI} plus 
a few young isochrones specifically computed for this project. 
In particular, we adopted isochrones based on evolutionary tracks 
accounting for mass-loss ($\eta$=0.4), neglecting both convective
overshooting during the core H-burning phases and atomic 
diffusion (Pietrinferni et al.\ 2004, 2006).  
We have assumed a true distance 
modulus of $\mu$=24.60 and a reddening E($F555W$-$F814W$)=1.16 mag. 
Data plotted in the bottom left panel of Fig.~1 show that young 
scaled-solar isochrones (red lines) at fixed metal and helium content 
($[M/H]$=--0.66, $Y$=0.251) and ages ranging from 
6 Myr ($M(Turn-Off [TO])/M_\odot$$\sim$ 30.5) to 
100 Myr ($M(TO)/M_\odot$$\sim$ 4.3) properly fit young
MS stars in IC10. The same conclusion applies to the fit of RGB stars 
with the old $\alpha$-enhanced isochrone (t$\sim$14 Gyr). 
Note that the global metallicity of this isochrone is 
$[M/H]$=--0.66 with $[Fe/H]$=--1.01 and $[\alpha/Fe]\sim$0.4 
(Pietrinferni et al.\ 2006). 
The data plotted in the bottom panel of Fig.~1 were selected 
according to photometric error ($\sigma_{F814W}$=$\sigma_{F555W}$$\le$ 0.1), 
{\em separation} ($sep$ $\ge$ 4) and {\em sharpness} ($|sha|$$\le$0.3). 
The {\em separation index} quantifies the degree of crowding 
(Stetson et al.\ 2003). The current value corresponds to stars 
that have required a correction of less than a few percent for 
light contributed by known neighbours. The {\em sharpness index}
quantifies the similarity between the shape of the measured objects 
and of the Point-Spread-Function (PSF). It is the quadrature difference 
between the one-sigma-half-width of the measured object and the 
one-sigma-half-width of the core of the PSF (Stetson \& Harris 1988).  
The bottom right panel of Fig.~1  shows that the CMD of the 
external regions is deeper, since these regions are less affected by 
crowding. The comparison with the central regions indicates a significant 
%
change in age distribution,  
and indeed, massive MS stars almost disappear when moving 
toward the external regions. The MS stars are well fit by isochrones 
with ages ranging from 12 Myr ($M(TO)/M_\odot$$\sim$ 14.2) to 
350 Myr ($M(TO)/M_\odot$$\sim$ 2.5). These findings and the continuous 
stellar distribution along the MS indicate that IC10 experienced ongoing 
star formation during $\approx$ the last half Gyr. Fig.~1 
also indicates that IC10's stellar populations show a spread in metal 
content. The width in color of RGB stars is well fit by isochrones 
with a single age (14 Gyr) and metal contents ranging from   
$[M/H]\sim-0.96$ ($Y=$0.248, green line) to  $[M/H]\sim-0.35$ 
($Y=$0.256, blue line).
%
The above comparison between theory and observations has to be 
considered as a preliminary guideline. These estimates of age 
and metal content are affected by empirical (distance modulus, 
reddening, photometric zero-points) and theoretical (mixing length, 
color-temperature relations) uncertainties. Firm constraints on these 
parameters require deep and accurate photometry down to the TO of the 
old population.      
Note that the possible presence of differential reddening 
amounting to $\pm$10\%  would not account for the observed spread in 
color (see the arrows in the right panel of Fig.~1 and Sanna et al.\ 2008).   
%
However, IC10 has an extended HI envelope, a large number of 
HII regions (Hidalgo-Gamez 2005) and molecular clouds 
(Leroy et al.\ 2006). This means that internal spatial variations 
of the reddening are quite probable. To partially overcome this problem, 
we adopted as representative of the IC10 stellar content the stars located 
in a small external region (E, see the red polygon in the top panel 
of Fig.~1).   
%

\begin{figure}[!ht]
\begin{center}
\label{fig2}
\includegraphics[height=0.50\textheight,width=0.50\textwidth]{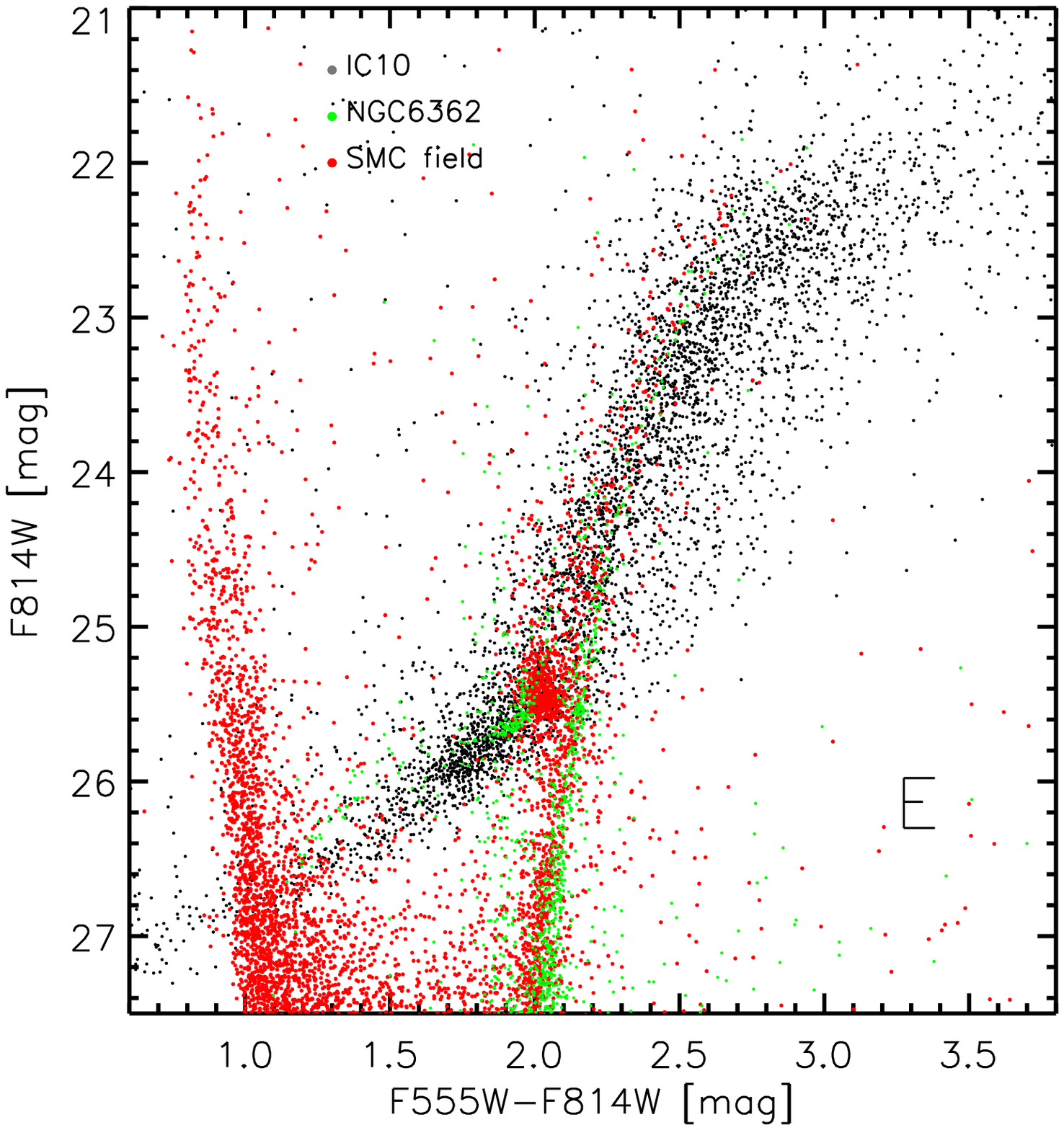}
\caption{$F814W$, $F555W$-$F814W$ CMD of IC10 external regions
(black dots) compared with an SMC field (red dots) and with the 
GC NGC~6362 (green dots). The number of IC10 stars plotted in this 
figure is similar to the number of stars in the SMC field. See text 
and Table~1 for more details concerning the true distance moduli 
and the reddenings adopted for these systems. 
}
\end{center}
\end{figure}

To further characterize the stellar content of IC10 we also adopted 
empirical tracers. Fig.~2 shows the comparison between field {\em E)} 
of IC10 and the ACS photometry of an SMC field provided by 
Sabbi et al.\ (2007; red dots). 
Note that in this figure we plotted a number of IC10 stars, randomly 
selected, similar to the number of stars present in the SMC field. 
The SMC sample was plotted by assuming 
for the SMC a true distance modulus of $\mu=18.9$ and a reddening of 
$E(B-V)=0.08$ mag. The ages of the main stellar components in this SMC field 
range from a few tens of Myr (bright MS) to a few Gyr (red clump, RC). 
To investigate the possible presence of an old stellar population 
we also compared IC10 to the globular cluster NGC~6362. The $V,I$-band photometry 
(Stetson 2000) for this cluster was transformed into the ACS photometric 
system using the transformations by Sirianni et al.\ (2005). NGC~6362 
is an almost metal-rich cluster ($[Fe/H]=-1.04\pm0.06$, $[M/H]\sim$-0.75, 
see Table~1) and its Horizontal Branch (HB) morphology is characterized 
by both red and blue stars (Brocato et al.\ 1999). 
Data plotted in this figure, in particular in the helium burning 
region (i.e., RC and red HB stars, 25.8$\lesssim$$F814W$$\lesssim$25.2, 
1.7$\lesssim$$F555W-F814W$$\lesssim$2.1), indicate that IC10 hosts both  
intermediate-mass stars (RC) and candidate old low-mass (red HB) stars. 
To further constrain these key points Fig.~3 shows the 
comparison of the same field with three GCs with different metal contents 
($[M/H]\sim$--0.47, 47 Tuc; --0.85, NGC~2808; --0.93, NGC~1851, see references 
listed in Table~1) and HB morphologies (only red HB stars, 47 Tuc; red, blue HB 
and RR Lyrae stars, NGC~1851; red, blue HB and extreme HB stars, NGC~2808). 
Note that in this figure we plotted a number of IC10 stars, randomly 
selected,  similar to the number of stars present in the GC NGC~2808. 
Data in Fig.~3 further support the evidence (see Fig.~1) that RGs in 
IC10 cover at least one-half dex in metal content 
(--0.4$\lesssim$$[M/H]$$\lesssim$--1). Moreover, a fraction of the stars 
located near $F814W$$\approx$26 and 
1.5$\lesssim$$F555W-F814W$$\lesssim$1.8 appear to be candidate 
RR Lyrae stars.  Finally, we note that the comparison with empirical 
calibrators indicates that both metal-intermediate and metal-rich 
candidate RGB bump stars (see Table~1) 
%
should be systematically fainter than our current limiting
magnitude at the color typical of RGB bump stars
($F555W-F814W$$\approx$2.1--2.2 mag, see the arrows in Fig.~3). 

\begin{figure}[!ht]
\begin{center}
\label{fig3}
\includegraphics[height=0.50\textheight,width=0.50\textwidth]{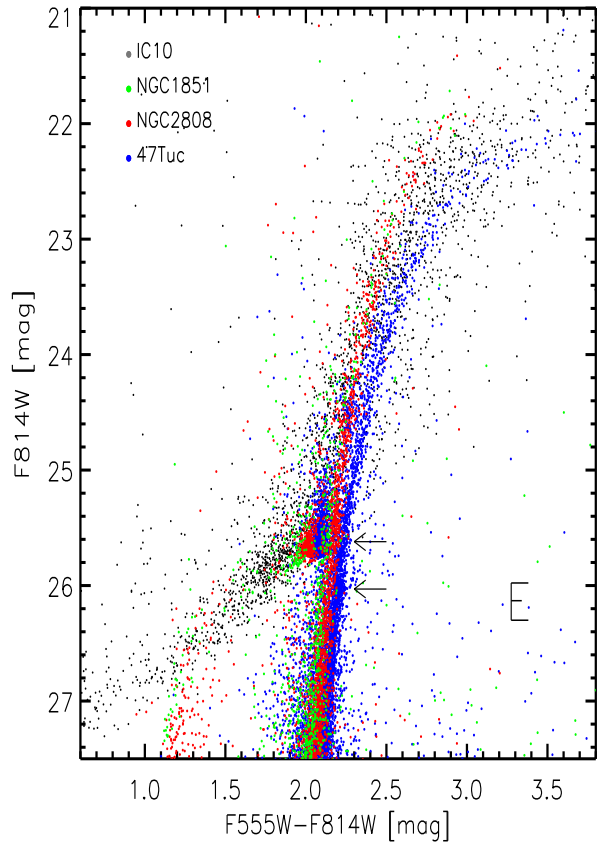}
\caption{Same as Fig.~2, but the comparison is performed with three 
GCs: NGC~1851 (green dots), NGC~2808 (red dots) and 47Tuc (blue dots).
The number of IC10 stars plotted in this figure is similar to the 
number of stars in NGC~2808. 
The true distance moduli and reddenings adopted for these systems 
are listed in Table~1. From top to bottom the arrows display the 
position of the RGB bumps for NGC~1851 and 47Tuc.  
}
\end{center}
\end{figure}
 
To investigate the properties of He-burning stars we adopted different 
sets of evolutionary models constructed assuming both old and 
intermediate-age progenitors. The top panel of Fig.~4 shows 
$\alpha$-enhanced Zero-Age-Horizontal-Branch (ZAHB, solid line) 
models together with the He-exhaustion locus (dashed, 10\% of central 
He still available) at fixed metal and helium content ($[M/H]$=--0.66, $Y$=0.251)
for an old ($sim14$ Gyr) progenitor ($M_{pr}$=0.80 $M_\odot$). 
To cover the age range of IC10 stars we also adopted core He-burning (solid) 
and He-exhaustion (dashed, 10\% of He left) models for a set of intermediate-age 
(160 Myr $\le$$t$$\le$ 6.6 Gyr) progenitors 
(1.0$\le$$M_{pr}$$\le$3.5 $M_\odot$).  
The evolutionary properties of He-burning, low-to-intermediate mass stars 
have been thoroughly investigated in the literature (Sweigart, Greggio \&  
Renzini 1990; Castellani et al.\ 2000; Pietrinferni et al.\ 2004,2006; 
Bertelli et al.\ 2008). Here we note that the ratio between 
He- and H-burning lifetimes is quite constant ($t_{He}/t_H$=0.006) 
when moving from $M(HB)$=0.60 to 0.80 $M_\odot$ (old progenitor).
On the other hand, the same ratio changes from $t_{He}/t_H$=0.11 
($M(RC)$=1.83$M_\odot$) to 0.39 ($M(RC)$=2.18$M_\odot$) and to 0.34 
($M(RC)$=2.78$M_\odot$)\footnote{The use of He lifetimes at He-exhaustion, 
i.e., no He left in the core, changes the quoted ratios by a few 
thousandths and a few hundredths for old and intermediate-age 
progenitors, respectively.} for scaled-solar, intermediate-mass progenitors 
(see top panel of Fig.~4). This stark difference is caused by the 
fact that when moving from 0.8 to 2.2 $M/M_\odot$ the core He-ignition 
takes place in structures that are less and less affected by electron 
degeneracy. This means that the He core mass at He-ignition, and in 
turn the luminosity during core He-burning, decreases from 
$M^c_{He}/M_\odot$=0.485 ($M_{pr}$=0.80 $M_\odot$, $M_{F814W}$=--0.35 mag, 
$t_H\sim$14 Gyr) to $M^c_{He}/M_\odot$=0.467 ($M_{pr}$=2.2 $M_\odot$, 
$M_{F814W}$=--0.05 mag, $t_H$=750 Myr). 
More massive structures are characterized by a convective core during 
H-burning phases and a further increase in stellar mass causes a steady 
increase in the He core mass and in luminosity  ($M_{pr}$=2.80 $M_\odot$, 
$M^c_{He}$=0.370 $M_\odot$, $M_{F814W}$=--1.02 mag, $t_H$=280 Myr). 
The above difference implies that the expected star count ratio between MS 
and He-burning structures increases by 1--2 orders of magnitude when moving 
from old to intermediate-mass stars.   

Therefore, we decided to perform a more detailed comparison between theory 
and observations. We selected stars in the external regions using severe 
criteria ($\sigma_{F814W}$=$\sigma_{F555W}$$\le$ 0.07 mag, $sep$ $\ge$ 6, 
$|sha|$$\le$0.2). Data plotted in the bottom panel of Fig.~4 show that 
the well defined overdensity centered on  $F814W$$\sim$25.72 and 
$F555W-F814W$$\sim$1.90 agrees quite well with the expected position 
of old, red HB stars and of intermediate-age red clump stars. 
The solid lines display the ZAHBs for two different metal contents 
($[M/H]$=--0.96, $Y=$0.248; $[M/H]$=--0.35, $Y=0.256$) and for old 
and intermediate-age progenitors. The stellar masses and the ages of 
the progenitors are quite similar to the models plotted in the 
top panel. Note that the 
spread in magnitude and color of the He-burning region is larger 
than the typical photometric errors (see error bars). The range in 
color covered by the stellar overdensity is also systematically 
bluer and larger than the color range covered by typical 
low-mass RGB bump stars. The RGB bump in a metal-rich (47Tuc) 
and in a metal-intermediate (NGC~1851) globular cluster is, 
indeed, fainter and redder (see the arrows in Fig.~3).     
This comparison also supports the hypothesis that the stars with 
$F814W$$\sim$26 and $F555W-F814W$$\sim$1.7 are candidate 
RR Lyrae stars. Note that corrections for completeness of the current 
photometry would go in the direction of increasing the number of candidate 
old HB stars.   

The photometric accuracy in the region around the peak does not 
allow us to distinguish RC from old HB stars. However the occurrence of 
warm HB stars, once confirmed by independent experiments, will provide 
a robust identification of the so-called {\em Baade's red sheet}, 
i.e., evidence for an old stellar population (Baade 1963) 
in a starburst galaxy. Current circumstantial evidence is, indeed, 
based on intermediate-age (RC) He-burning stars (Aparicio et al.\ 1997; 
Schulte-Ladbeck et al. 1998). Moreover, the 
identification of massive MS stars, old (HB), and intermediate 
age (RC) helium burning stars indicates that IC10 underwent several 
star formation episodes during its life. A CMD a couple 
of magnitudes deeper and with a stronger temperature sensitivity 
could provide firm constraints on whether the star formation activity 
of this interesting system has been continuous or sporadic.

\begin{figure}[!ht]
\begin{center}
\label{fig4}
\includegraphics[height=0.60\textheight,width=0.50\textwidth]{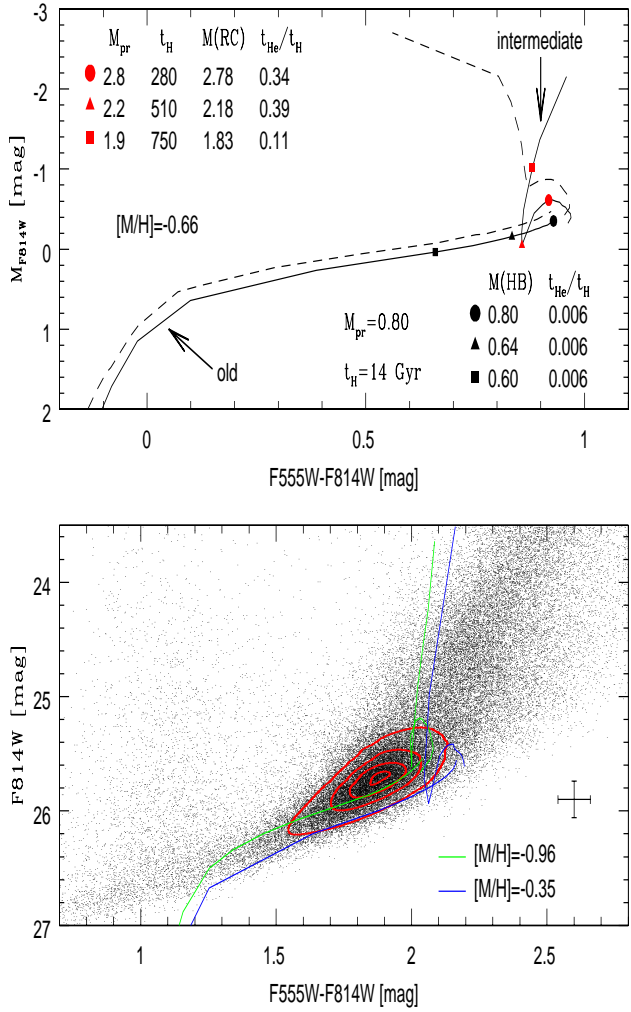}
\caption{Top -- Color-Magnitude diagram for predicted He-burning structures
at fixed global metallicity ($[M/H]$=--0.66). 
The fainter solid and dashed lines show the $\alpha$-enhanced ZAHB and the 
He-exhaustion (10\% of He core still available) for an old progenitor 
($t_H$=14 Gyr, $M_{pr}$=0.80 $M_\odot$). The brighter solid and dashed lines 
show the core He-burning and the He exhaustion for intermediate-age 
progenitors. The mass of the progenitors, the mass at core He-burning 
and the ratio between He and H lifetimes for selected structures are 
labeled and marked with black (old) and red (intermediate-age) symbols, 
respectively.  Bottom -- The red solid polygons display the 
35, 60, 80 and 97\% isodensity levels. 
Fainter  blue and green lines show the ZAHBs for old, low-mass structures 
with different chemical compositions. The almost vertical green and 
blue lines display core He-burning structures with the same compositions, 
but for intermediate-mass structures.}
\end{center}
\end{figure}

\acknowledgments
It is a pleasure to thank P. Popesso for several interesting discussions
concerning dwarf galaxies. We also thank an anonymous referee for his/her 
pertinent comments and detailed suggestions that helped us to improve the 
content and the readability of the manuscript. This project was partially 
supported by the grant Monte dei Paschi di Siena (P.I.: S. Degl'Innocenti), 
PRIN-INAF2007 (P.I.: M. Bellazzini), PRIN-MIUR2007 (P.I.: G. Piotto).


\begin{deluxetable}{llllcccc}
\scriptsize
\tablewidth{0pt}                       
\tablecaption{Intrinsic parameters of the GCs adopted as empirical calibrators.}
\tablehead{
\colhead{ID}&
\colhead{$\mu$\tablenotemark{a}}&
\colhead{$E(B-V)$\tablenotemark{b}}&
\colhead{$[Fe/H]_{spe}^c$}&
\colhead{$[Fe/H]_{ZW}^d$}&
\colhead{$[Fe/H]_{CG}^d$}&
\colhead{$[M/H]$\tablenotemark{e}}&
\colhead{RGB bump\tablenotemark{f}}
}
\startdata
47Tuc    & $13.32\pm0.09$\tablenotemark{g} & $0.04\pm0.02$\tablenotemark{h}  & $-0.76\pm0.04$\tablenotemark{i}  &
$-0.71\pm0.05$  & $-0.78\pm0.02$  & $-0.47$ & $13.49 \pm 0.10$\\
NGC6362   & $14.43\pm0.05$\tablenotemark{j} & $0.08\pm0.02$\tablenotemark{j}  & $-1.04\pm0.06$\tablenotemark{k}  & 
$-1.18\pm0.06$  & $-0.99\pm0.03$  & $-0.75$ &  $14.45 \pm 0.10$ \\
NGC2808   & $15.23\pm0.10$\tablenotemark{l} & $0.18\pm0.01$\tablenotemark{m}  & $-1.14\pm0.10$\tablenotemark{n}  & 
$-1.36\pm0.05$  & $-1.11\pm0.03$  & $-0.85$ &  $15.10 \pm 0.10$ \\
NGC1851   & $15.44\pm0.20$\tablenotemark{o} & $0.02\pm0.02$\tablenotemark{o}  & $-1.22\pm0.03$\tablenotemark{p}  &
$-1.23\pm0.11$  & $-1.03\pm0.06$  & $-0.93$ & $15.16 \pm 0.10$\\
\enddata
\tablenotetext{a}{Cluster true distance modulus (mag).}  
\tablenotetext{b}{Cluster reddening (mag).}  
\tablenotetext{c}{High-resolution spectroscopic iron abundances.}  
\tablenotetext{d}{Iron abundances based on Ca triplet measurements provided by 
Rutledge et al.\ (1997) in the Zinn \& West (1984) and in the Carretta \& 
Gratton (1997) metallicity scale.}
\tablenotetext{e}{Global metallicity based on spectroscopic iron abundances 
and assuming $[\alpha/Fe]=0.4$ (Salaris et al.\ 1993).}
\tablenotetext{f}{The Johnson-Cousins magnitudes of the RGB bump provided by
Di Cecco et al.\ (2009, in preparation) were transformed into the  ACS 
$F814W$-band (VEGAMAG) following Sirianni et al.\ (2005).}
\tablenotetext{g}{Bono et al.\ (2008). $^h$Salaris et al.\ (2007).  
$^i$Koch et al.\ (2008).  $^j$Brocato et al.\ (1999). $^k$Gratton (1987). 
$^l$Castellani et al.\ (2006). $^m$Bedin et al.\ (2000).  $^n$Carretta (2006).  
$^o$Saviane et al.\ (1998).  $^p$Yong et al.\ (2009).}
\end{deluxetable}

\end{document}